\documentclass[twocolumn,showpacs,preprintnumbers,prb]{revtex4}

\usepackage{graphicx}
\usepackage{dcolumn}
\usepackage{bm}

\begin{document}

\title{Pressure suppression of unconventional charge-density-wave 
state in PrRu$_4$P$_{12}$ \\
studied by optical conductivity}

\author{H.~Okamura}\altaffiliation[Electronic address: ]{okamura@kobe-u.ac.jp}
\author{N. Ohta}
\author{A. Takigawa}
\author{I. Matsutori}
\author{K. Shoji}
\author{K. Miyata}
\author{M. Matsunami}\altaffiliation[Present address: ]
{UVSOR Facility, Institute for Molecular Science, Okazaki 444-8585, Japan}
\author{H. Sugawara}
\affiliation{Department of Physics, Graduate School of Science, 
Kobe University, Kobe 657-8501, Japan} 

\author{C. Sekine}
\author{I. Shirotani}
\affiliation{Department of Electrical and Electronics Engineering, 
Muroran Institute of Technology, Muroran 050-8585, Japan} 

\author{H. Sato}
\affiliation{Graduate School of Science, Tokyo Metropolitan University, 
Hachioji 192-0397, Japan} 

\author{T. Moriwaki}
\author{Y. Ikemoto}
\affiliation{Japan Synchrotron Radiation Research Institute and SPring-8, 
Sayo 679-5198, Japan} 

\author{Z. Liu}
\affiliation{Geophysical Laboratory, Carnegie Institution of Washington, 
Washington DC 20015, USA} 

\author{G. L. Carr}
\affiliation{National Synchrotron Light Source, Brookhaven 
National Laboratory, Upton, New York 11973, USA}

\date{\today}

\begin{abstract}
Optical conductivity [$\sigma(\omega)$] of PrRu$_4$P$_{12}$ 
has been studied under high pressure to 14~GPa, at low 
temperatures to 8~K, and at photon energies 12~meV-1.1~eV.  
  The energy gap in 
$\sigma(\omega)$ at ambient pressure, caused by a 
metal-insulator transition due to an unconventional 
charge-density-wave formation at 63~K, is gradually 
filled in with increasing pressure to 10~GPa.       
At 14~GPa and below 30~K, $\sigma(\omega)$ exhibits a 
pronounced Drude-type component due to free carriers.  
This indicates that the initial insulating ground state 
at zero pressure has been turned into a metallic one 
at 14~GPa.     
This is consistent with a previous resistivity study 
under pressure, where the resistivity rapidly decreased 
with cooling below 30~K at 14~GPa.    
The evolution of electronic structure with pressure is 
discussed in terms of the hybridization between the 
4$f$ and conduction electrons.    
\end{abstract}

\pacs{ }

\maketitle
\section{Introduction}
The metal-insulator (MI) transition in PrRu$_4$P$_{12}$ has 
attracted a great amount of attention since its discovery.\cite{sekine}   
PrRu$_4$P$_{12}$ has the filled skutterudite structure, where 
a Pr atom is contained in a cage-shaped P$_{12}$ molecule.    
Due to the large coordination number, the hybridization between 
the Pr 4$f$ electrons and the conduction electrons, the latter 
of which are mainly P 3$p$ electrons, is expected 
to be stronger than in many other Pr compounds.   
This property has made PrRu$_4$P$_{12}$ an attractive stage 
to study the interaction between 4$f^2$ configuration with the 
conduction electrons.  
The MI transition is observed at $T_{\rm MI}$=63~K, 
below which the electrical resistivity ($\rho$) 
shows a sharp 
upturn.\cite{sekine,saha}    
With a single crystal sample, $\rho$ increased by a factor 
of $\sim$ 400 from $T_{\rm MI}$ to 2~K.\cite{saha}   
From the analysis of $\rho(T)$ data, the magnitude of the 
energy gap was estimated to be $\sim$ 7~meV.\cite{sekine}   
In addition, 
the development of an energy gap was clearly observed in 
the measured optical conductivity [$\sigma(\omega)$] 
spectrum.\cite{matunami}

In discussing the mechanism for this transition, 
it was pointed out from band calculations that the Fermi 
surface (FS) of PrRu$_4$P$_{12}$ should have a cube-like 
shape, with a strong tendency for three dimensional (3D) 
nesting.\cite{harima1,harima2}  
In fact, superlattice formation below $T_{\rm MI}$ was 
observed by diffraction experiments.\cite{lee}    
However, the non-magnetic reference LaRu$_4$P$_{12}$ did 
not show a similar transition, although a very similar 
FS was expected.\cite{harima1,harima2}   
This fact, together with the neutron\cite{iwasa1,iwasa2} 
and Raman\cite{ogita} experiments, strongly suggested 
the importance of conduction 
($c$)-$f$ electron hybridization in the MI transition.    
With theoretical considerations,\cite{takimoto,shiina1,shiina2} 
it is currently believed that a novel charge density wave 
(CDW) formation is responsible for the 
MI transition.   Unlike the conventional CDW where 
the spatial modulation is provided by strong 
electron-lattice coupling, in the CDW of PrRu$_4$P$_{12}$ 
the modulation is provided by the formation of two Pr 
sublattices, where the crystal field-split 4$f$ level 
scheme in one sublattice is different from that in the 
other.\cite{iwasa1,iwasa2,ogita}    
The microscopic nature of the $c$-$f$ hybridized 
electronic structure deserves further studies.

The MI transition in PrRu$_4$P$_{12}$ 
has been found to be sensitive to 
external pressure ($P$).\cite{shirotani,miyake1,miyake2}   
With increasing $P$ up to 8~GPa, the increase of $\rho$ 
below $T_{\rm MI}$ becomes gradually smaller.\cite{shirotani}  
At 12~GPa, a small upturn of $\rho$ is still seen below 
$\sim$ 60~K, but a rapid decrease of $\rho$ follows 
with further cooling below 30~K.\cite{miyake2}    
An external pressure is expected to reduce both the Pr-P 
and Pr-Pr distances, and is hence expected to affect the 
$c$-$f$ hybridization and the CDW state.    
It is therefore quite interesting to probe the 
evolution of $\sigma(\omega)$ with $P$.

In this work, we have studied $\sigma(\omega)$ of 
PrRu$_4$P$_{12}$ under high pressure to 14~GPa and at 
low temperatures ($T$) to 8~K.     The obtained $\sigma(\omega)$ 
spectra show that the well developed energy gap at 
ambient pressure and at low $T$ is gradually filled 
in with increasing pressure up to 10~GPa.    
At 14~GPa and at 8~K, the energy gap is almost 
absent and $\sigma(\omega)$ shows a pronounced Drude-type 
spectral shape due to free carriers.   This result 
indicates that the electronic structure of PrRu$_4$P$_{12}$ 
at this $P$ and $T$ range is indeed metallic.

\section{Experimental} 
$\sigma(\omega)$ of a sample at high pressure was obtained 
from the analysis of a reflectance spectrum, $R(\omega)$, 
measured with a diamond anvil cell 
(DAC).\cite{ybs,airapt,CeRu4Sb12,heavy,Yb-metal}    
A pair of diamond anvils with 0.6~mm culet diameter and a 
metal gasket (stainless steel 301) 
were used to seal the sample, pressure transmitting medium, 
and small ruby pieces used to monitor the pressure via its 
fluorescence.\cite{airapt}    A flat, as-grown surface of a 
single crystal sample\cite{saha} was closely attached on 
the culet surface of the anvil, which was of type 
IIa with low density of impurities.    
The pressure-transmitting medium used was glycerin.\cite{medium}    
A gold film was also placed between the gasket and anvil 
as a reference of reflectance.   
Due to the small size of the sample and the restricted 
sample space in a DAC, synchrotron radiation was used as 
a highly oriented and bright infrared (IR) source.  The 
high pressure studies were made mainly at the beam line 
BL43IR of SPring-8\cite{micro1,micro2} at photon 
energies between 20~meV and 1.1~eV.  Above 1.1~eV, the 
reflectance at ambient temperature and 
pressure\cite{matunami} was used.   
Additional measurements at lower photon energies to 12~meV 
were also made at the beam line U2A of National Synchrotron 
Light Source.\cite{u2a}   
In the experiments, the pressure was increased at room 
$T$ to a desired value, and then $T$ was lowered while 
keeping the desired pressure.   
A liquid He continuous-flow cryostat was used to cool 
the DAC.    More details of the high pressure IR 
experiments can be found elsewhere.\cite{ybs,airapt,CeRu4Sb12}   
The $R(\omega)$ measurement without using a DAC was 
previously described.\cite{my_review,ybb12,ybal3}

\section{Results and Discussion}  
\subsection{$R(\omega)$ data at high pressure}    
Figure~1 summarizes $R(\omega)$ 
data measured at $P$=0, 8, 10, and 14~GPa.      
%
\begin{figure}[t]
\begin{center}
\includegraphics[width=0.45\textwidth,clip]{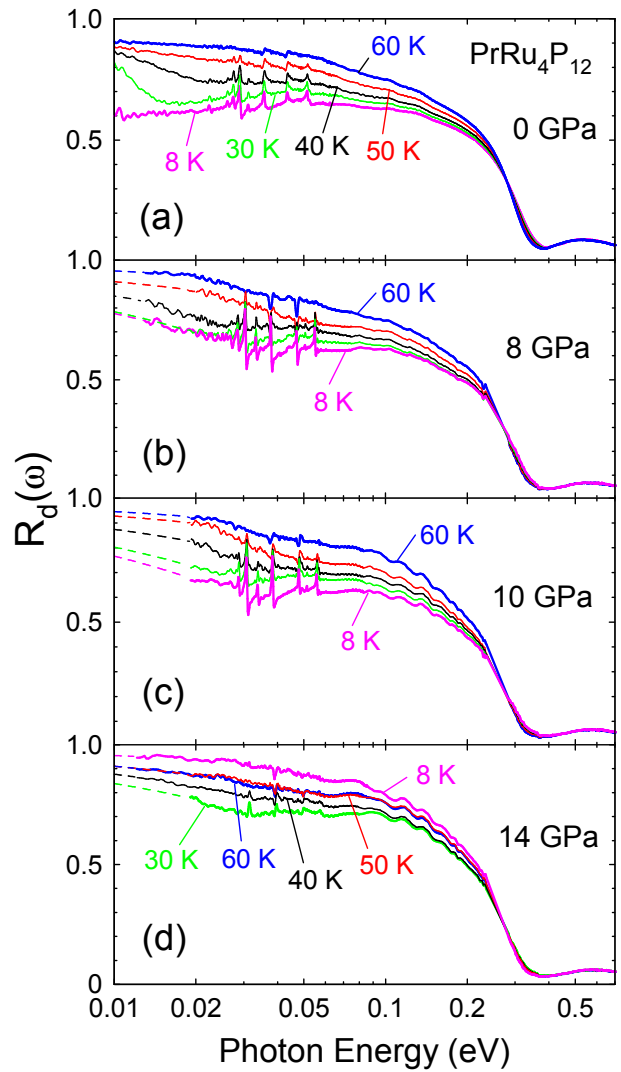}
\caption{(color online) 
Optical reflectance relative to diamond 
[$R_{\rm d}(\omega)$] of PrRu$_4$P$_{12}$ at several 
pressures and temperatures.  The broken curves 
indicate the extrapolations discussed in the text.  
The spectra in (a) 
were derived from previous data measured at 
zero external pressure,\cite{matunami} and those 
in (b)-(d) were obtained in this work.  
} 
\end{center}
\end{figure}
%
These spectra were obtained with the following procedures.   
The reflectance in vacuum, $R_0(\omega)$, were first 
measured without using a DAC over a wide 
photon energy range between 6~meV and 30~eV.\cite{matunami}    
The usual Kramers-Kronig (KK) transform\cite{dressel,wooten} 
was applied to the $R_0(\omega)$ to obtain the complex 
refractive index $\hat{n}(\omega)$.    $\hat{n}(\omega)$ was 
then used to derive the reflectance spectrum, 
$R_{\rm d}(\omega)$, that was {\it expected} in a DAC, 
taking into account the refractive index of 
diamond (2.4) as previously 
discussed.\cite{ybs,airapt,CeRu4Sb12,heavy}  
The $R_{\rm d}(\omega)$ data at $P$=0 in Fig.~1(a) 
are the results of this analysis.  
In the high pressure studies, the relative changes of 
$R_{\rm d}(\omega)$ with varying $P$ and $T$ were 
measured with DAC, and were then multiplied by the expected 
$R_d(\omega)$ at $P$=0 discussed above, to obtain the 
$R_{\rm d}(\omega)$ spectra at high pressure.     
This procedure was taken since it was technically difficult 
to accurately determine the absolute value of $R_{\rm d}(\omega)$ 
in a DAC, due to strong diffraction of the long wavelength 
IR radiation off the small sample.\cite{footnote3} 
For technical reason, lower-energy measurements to 12~meV 
were made only at $T$=8, 40, and 60~K and at 
$P$= 8 and 14~GPa.    
For other values of $T$ and $P$, the measurements 
were made to 20~meV only.  
In the $P$=0 data of Fig.~1(a), $R_{\rm d}(\omega)$ 
progressively decreases with cooling below 60~K.   This is 
due to the development of energy gap upon the MI 
transition.\cite{matunami}   At $P$= 8~GPa, a decrease 
of $R_{\rm d}(\omega)$ with cooling is still observed, 
but is much less pronounced below 40~K and 20~meV.    
At 10~GPa, the data are very similar to those at 8~GPa.  
At 14~GPa, however, the decrease of $R_{\rm d}(\omega)$ with 
cooling from 60 to 30~K is much smaller than those seen 
at lower $P$, and then $R_{\rm d}(\omega)$ considerably 
increases with cooling from 30 to 8~K.\cite{irizawa}   
It is quite clear that the $T$ dependence at 14~GPa 
is qualitatively very different from those at lower $P$.

\subsection{$\sigma(\omega)$ data at high pressure}
To derive $\sigma(\omega)$ from the $R_{\rm d}(\omega)$ data 
at high pressure measured with DAC, a modified KK analysis 
was performed on $R_{\rm d}(\omega)$ taking into account the 
effects of sample/diamond interface as previously described 
in detail.\cite{KK-dia}     
To extrapolate the measured $R_{\rm d}(\omega)$ spectra at the 
lower energy end, a Hagen-Rubens function\cite{dressel,wooten} 
was used.   At the higher energy side, 
$R_{\rm d}(\omega)$ spectra were cut off at 2.2~eV, above which 
they were extrapolated with a function of the form $\omega^{-4}$.    
The cutoff energy of 2.2~eV was chosen so that the 
$\sigma(\omega)$ obtained from the modified KK analysis of 
the expected $R_{\rm d}(\omega)$ at ambient [Fig.~1(a)] agreed 
with that obtained from the usual KK analysis of $R_0(\omega)$ 
up to 30~eV measured without DAC.\cite{KK-dia}

Figure~2 summarizes the obtained $\sigma(\omega)$ spectra at 
high pressure.     
%
\begin{figure}[t]
\begin{center}
\includegraphics[width=0.4\textwidth,clip]{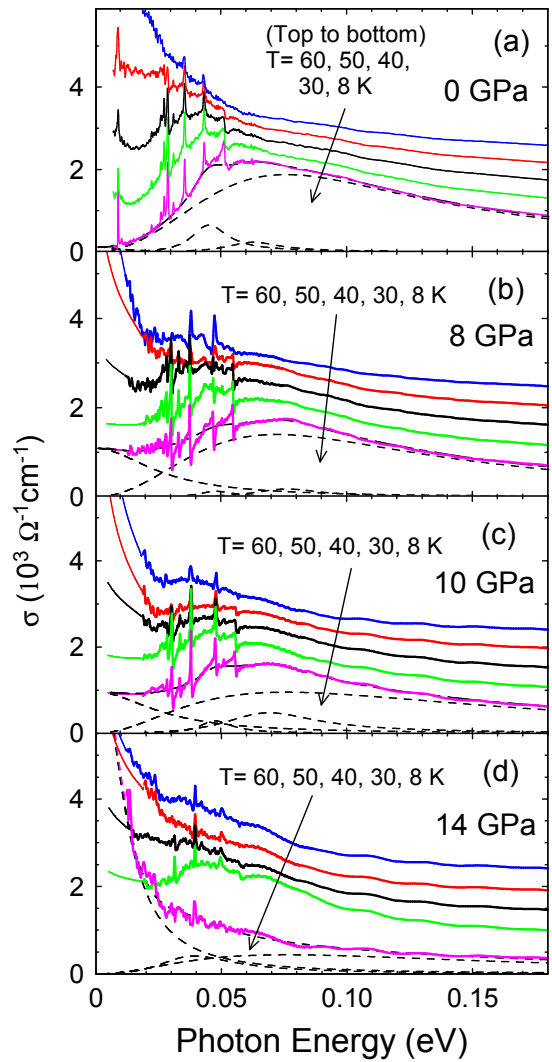}
\caption{(color online) 
Optical conductivity ($\sigma$) of PrRu$_4$P$_{12}$ 
at different pressures and temperatures.   
For clarity, each spectrum is vertically offset by 
500~$\Omega^{-1}$cm$^{-1}$.   The thin curves correspond 
to the extrapolated portion of $R_{\rm d}(\omega)$.  
The broken curves indicate the results of Drude-Lorentz 
spectral fitting for the 8~K data discussed in the text.  
} 
\end{center}
\end{figure}
(The $P$=0 data were obtained from $R_0(\omega)$ with the 
usual KK analysis.)
At $P$=0, $\sigma(\omega)$ shows a rapid 
decrease and depletion of spectral weight below 40~meV with 
cooling, and at 8~K, it shows a well developed energy gap as 
described previously.\cite{matunami}   (Note that each 
spectrum in Fig.~2 is vertically offset for clarity.)  
The onset of $\sigma(\omega)$ suggests a gap magnitude of 
about 10~meV, 
which is close to the transport gap of 7~meV given by 
$\rho(T)$.\cite{sekine}    
In addition to the gap, a broad peak centered around 50~meV 
is observed, where sharp optical phonon peaks are superimposed.   
This broad peak should result from optical excitation of 
electrons across the energy gap, and will be referred to as 
the ``gap excitation peak''.  
At 8~GPa, $\sigma(\omega)$ below 50~meV still shows spectral 
depletion with cooling, similar to that at $P$=0.    
However, the depletion is now smaller than at $P$=0 and 
$\sigma(\omega)$ does not approach zero at the low-energy 
side.  Namely, $\sigma(\omega)$ at 8~GPa has only an 
incomplete, partially filled energy gap even at low $T$.    
$\sigma(\omega)$ at 10~GPa shows similar $T$ evolutions to 
those at 8~GPa.  
$\sigma(\omega)$ at 14~GPa, however, shows very different $T$ 
evolutions.   Namely, $\sigma(\omega)$ 
at 14~GPa shows only small spectral depletion with cooling 
at $T$=30-50~K.  
At 8~K, then, $\sigma(\omega)$ shows a marked {\it increase} 
toward zero energy, which is clearly a free carrier component 
due to Drude-type response, and it does not show a clear 
energy gap any more.     
Furthermore, the phonon peaks in $\sigma(\omega)$ have become 
much weaker, which should result from an increased Coulomb 
screening of polarization due to a larger number of free 
carriers.     
Clearly, the electronic structure of PrRu$_4$P$_{12}$ at 
14~GPa and 8~K is metallic.

The observed $P$ and $T$ evolutions of $\sigma(\omega)$ show 
good correspondence with those of $\rho(T)$.\cite{shirotani,miyake1}    
Namely, 
the steep rise of $\rho(T)$ below $T_{\rm MI}$ at $P$=0 
is progressively suppressed with increasing $P$.\cite{shirotani}   
This is most likely due to the filling of energy gap with $P$ 
observed in $\sigma(\omega)$.    
On the other hand, the upturn of $\rho(T)$ is still 
observed near 60~K even at 8~GPa,\cite{shirotani} which is 
consistent with the result that spectral depletion in 
$\sigma(\omega)$ is still clearly observed at 8~GPa.    
At 14~GPa, $\rho(T)$ still shows small increase with cooling 
at 30-60~K range, but it then rapidly decreases with 
further cooling.     
These behaviors of $\rho(T)$ again correspond well to the 
$T$ variation of $\sigma(\omega)$.  Namely, $\sigma(\omega)$ 
still shows a small 
depletion below 40~meV with cooling from 50 to 30~K, but then 
shows a strong Drude-type component at 8~K.   
Namely, both $\rho(T)$ and $\sigma(\omega)$ at 14~GPa still 
seem to show a tendency toward gap formation at 30-60~K range, 
but then they show very metallic characteristics at 8~K.

Figure~3 shows the $P$ dependence of $\sigma(\omega)$ at 8 K.  
\begin{figure}[t]
\begin{center}
\includegraphics[width=0.38\textwidth,clip]{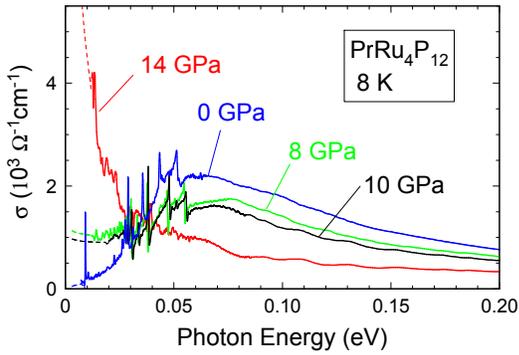}
\caption{(color online) 
Optical conductivity ($\sigma$) of PrRu$_4$P$_{12}$ 
at 8~K and at different pressures.   
The broken portion of the spectra indicates the 
extrapolated range.   } 
\end{center}
\end{figure}
With increasing $P$, the filling of energy gap and the 
transfer of spectral weight are evident.  
To analyze these evolutions more 
quantitatively, we have performed a Drude-Lorentz spectral 
fitting.\cite{dressel,wooten}   
The fitting model for a Lorentz oscillator is expressed as 
\begin{equation}
\sigma(\omega) = \frac{\omega_p^2}{4\pi}\frac{\gamma \omega^2}
{(\omega_0^2-\omega^2)^2 + \gamma^2 \omega^2},    
\end{equation}
where $\omega_p$ is the plasma frequency and $\gamma$ is the 
damping (scattering) frequency.  The Drude oscillator is 
given by setting $\omega_0$=0.  In the fitting, a linear 
combination of several oscillators was used, and the sharp 
phonon peaks were not taken into account.    
The results of the fitting are indicated by the broken curves 
in Fig.~2.  The peak position ($\omega_0$) and the spectral 
weight [$S$, which is the area of $\sigma(\omega)$] of the 
oscillators resulting from the fitting are plotted in Fig.~4.      
\begin{figure}
\begin{center}
\includegraphics[width=0.475\textwidth]{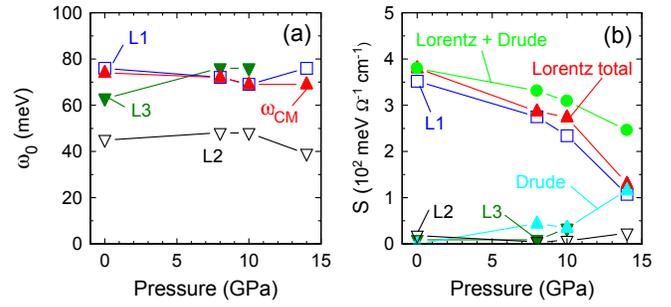}
\caption{
(a) The peak position ($\omega_0$) and (b) the spectral 
weight ($S$) of the Lorentz and Drude oscillators, which 
were obtained from the spectral fitting indicated in 
Fig.~2, plotted as a function of pressure.  
L1, L2, and L3 denote the three Lorentz oscillators 
in Fig.~2.  }
\end{center}
\end{figure}
%
As indicated in Fig.~2, three Lorentz oscillators (denoted as 
L1, L2, and L3 in Fig.~4) were necessary to systematically 
fit the gap excitation peak at different $P$.   
Lorentz oscillators were necessary even for the 14~GPa 
data, where the Drude component was dominant.   
To determine the position of the gap excitation peak 
and its shifts with $P$, we adapted a previously reported 
method used to analyze a CDW gap.\cite{RTe1,RTe2,RTe3,RTe4}    
Namely, the center-of-mass energy ($\omega_{\rm CM}$) of 
the three Lorentz oscillators used in the fitting, defined 
as 
$\omega_{\rm CM} = \sum_{i}{\omega_{0,i} S_i}/
\sum_{i}{S_i}$ where $i$ denotes L1, L2, and L3, 
was regarded as the peak position.      
As shown in Fig.~4, the obtained $\omega_{\rm CM}$ decreases 
only slightly with $P$ even at 14~GPa.      
In contrast, $S$ (Lorentz total) decreases much more 
strongly with $P$.  
In addition, the combined (Drude + Lorentz) $S$ 
also decreases with $P$, which suggests a spectral 
weight transfer to above the mid-infrared range.

%
The transfer of spectral weight in $\sigma(\omega)$ can 
also be discussed using the effective carrier density 
$N^\ast=n/m^\ast$, where $n$ and $m^\ast$ are the density 
and effective mass of the electrons.   
$N^\ast$ contributing to $\sigma(\omega)$ below $\omega$ 
is given by 
%
$N^\ast(\omega) = (2m_0/\pi e^2) \int_{0}^{\omega}
\sigma(\omega^\prime)d \omega^\prime$.    
%
Figure~5 plots $N^\ast(\omega)$ at various values of 
$P$ and $T$.   
%
\begin{figure}[t]
\begin{center}
\includegraphics[width=0.38\textwidth,clip]{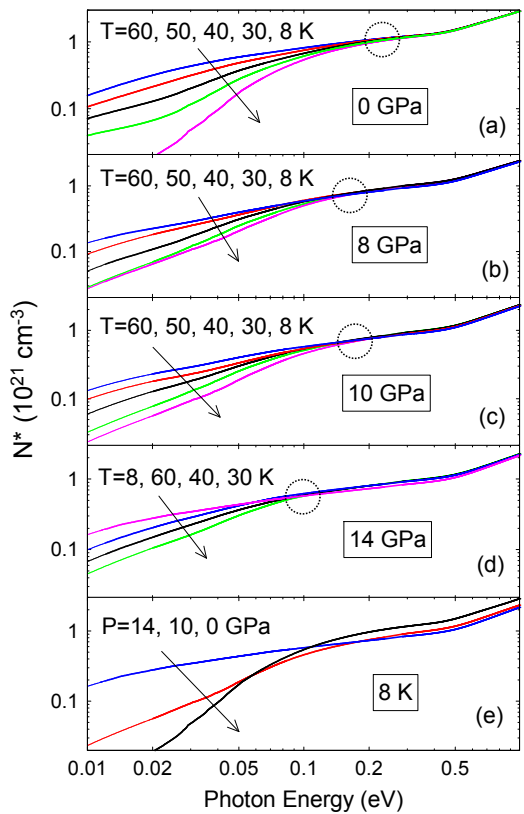}
\caption{(color online) 
Effective carrier density $N^\ast(\omega)$ of PrRu$_4$P$_{12}$.  
The graphs (a)-(d) show the $T$ dependences at a fixed 
pressure, while (e) shows the pressure dependence at 8~K.   
The circles indicate the energy ranges where the 
$N^\ast(\omega)$ spectra at different temperatures 
merge, as discussed in the text.   
In (d), the 50~K spectrum is not shown since it almost 
overlaps with the 60~K spectrum.  
} 
\end{center}
\end{figure}
At $P$=0 [Fig.~5(a)], the depletion of spectral weight 
with gap opening at low $T$ significantly decreases 
$N^\ast(\omega)$.  
In addition, the $N^\ast(\omega)$ spectra at different 
temperatures almost merge at 0.2-0.3~eV range.    
The energy range indicated by the circle in Fig.~5(a) 
shows such a merging.  
Note that the merging energy is a few times larger than 
$\omega_{\rm CM}$, and that the $N^\ast$ lost by 
the gap formation does not recover until such high 
energies.    This unusual result may also indicate an 
unconventional nature of this transition.    
With increasing $P$, the merging energy is seen to shift 
to lower energy, as indicated in 
Figs.~5(b)-5(d).  This shift shows that the energy range 
of spectral weight redistribution is also decreased as 
the energy gap is filled in with $P$.    
Note that, at 14~GPa, $N^\ast(\omega)$ still decreases 
from 50~K to 30~K, which corresponds to the above-mentioned, 
remaining tendency toward gap formation.  
Note also that the $N^\ast(\omega)$ value above the merging 
energy decreases at high $P$, which corresponds to the 
result in Fig.~4 that the total (Drude+Lorentz) 
$S$ decreases with $P$.

\subsection{Microscopic consideration on the energy 
gap and CDW state} 
As already mentioned, the magnitude of transport gap at 
$P$=0 given by $\rho(T)$ data is $E_g$=7~meV,\cite{sekine} 
and the onset of $\sigma(\omega)$ is at $\sim$ 10~meV.   
$E_g$ and the onset energy are close to each other, and 
they should correspond to a gap in the total density 
of states (DOS).  
In contrast, $\omega_{\rm CM} \sim$ 70~meV of 
the gap excitation peak is much larger.  Since the 
relationship between $E_{\rm g}$ and $\omega_{\rm CM}$ 
is unclear from the $\sigma(\omega)$ data only, 
we shall compare them with the results of microscopic 
theories.\cite{takimoto,shiina1,shiina2}   
In the theory of Ref.~\onlinecite{shiina2}, the Hamiltonian 
for PrRu$_4$P$_{12}$ contains the $c$-$f$ hybridization energy 
($J$), the bare crystal field (CF) at the Pr site ($\Delta_0$), 
and the intersite $f$-$f$ coupling energy ($K$) as the relevant 
parameters.   For simplicity, only two levels among the 
CF-split multiplets of Pr 4$f^2$, namely the singlet 
($\Gamma_1$) and triplet ($\Gamma_4$) levels, are assumed 
as the $f$ electron states.\cite{shiina2}   
This choice is made since they are the two lowest-lying CF 
levels observed experimentally.\cite{iwasa1,iwasa2,ogita}    
Above $T_{\rm C}$, all the Pr atoms are equivalent with 
singlet ground state.  
At $T$=0, due to the 3D nesting of FS and the $c$-$f$ 
hybridization, 
the ground state of this Hamiltonian is a CDW with 
two inequivalent Pr sublattices having triplet and 
singlet ground states.\cite{footnote4}

The phase diagram derived from the above model\cite{shiina2} 
is schematically shown in Fig.~6(a).  It is obtained from 
the mean-field (MF) solution to the Hamiltonian, and is 
expressed in terms of $T$ and the effective CF, 
$\Delta=\Delta_0 - J + \frac{3}{4}K$.   
\begin{figure}
\begin{center}
\includegraphics[width=0.47\textwidth,clip]{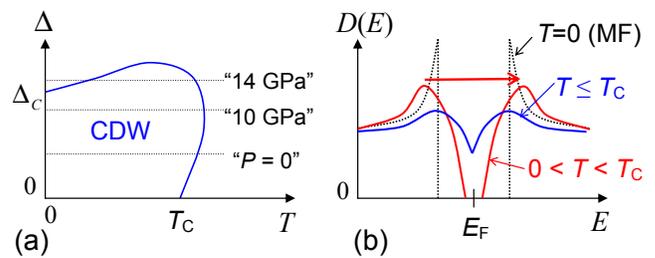}
\caption{(color online) 
Schematic illustrations of theoretical predictions in 
Ref.~\onlinecite{shiina2}.    (a) Phase diagram of PrRu$_4$P$_{12}$ 
in terms of $T$ and the effective crystal field, 
$\Delta = \Delta_0 - J +\frac{3}{4}K$. 
The three horizontal lines indicate the suggested positions 
of 0, 10, and 14~GPa.  
(b) Density of states as a function of energy [$D(E)$] around 
the Fermi level ($E_{\rm F}$) at $T$=0 (given by MF theory), 
at a low but finite $T$ ($0 < T < T_{\rm C}$), and just below 
$T_{\rm C}$ ($T \leq T_{\rm C}$).   
The horizontal (red) arrow indicates an optical transition for 
the gap excitation peak in $\sigma(\omega)$.   
The original $D(E)$ in Ref.~\onlinecite{shiina2} 
is asymmetric about $E_{\rm F}$, but here it is drawn 
symmetrically for simplicity. }
\end{center}
\end{figure}
Note that, for a range of $\Delta$, there is a 
possibility of successive, reentrant phase transitions 
with cooling.   
The DOS as a function of energy, $D(E)$, at finite $T$ is 
calculated with the coherent potential approximation, taking 
into account the thermal fluctuation of $f$ electron between 
the two CF levels.\cite{shiina2}    The predicted $D(E)$ is 
schematically shown in Fig.~6(b).      
It is seen that, with increasing $T$, the sharp DOS peaks 
associated with the $T$=0 (MF) gap are broadened, and the 
gap magnitude becomes much smaller, due to the thermal 
fluctuations.  
Clearly, the smaller gap in the total DOS should correspond 
to the transport gap and the onset of $\sigma(\omega)$.  
The optical transition between the 
remaining peaks in the DOS, indicated by the horizontal 
(red) arrow in Fig.~6(b), corresponds well to the observed 
gap excitation peak in $\sigma(\omega)$.  
In fact, the calculated $\sigma(\omega)$ shows a broad 
peak and an onset,\cite{shiina2} where the peak energy 
of the former is several times larger than the latter 
in good agreement with the measured $\sigma(\omega)$.      
The magnitude of CDW gap given by the MF theory is 
3.5~$k_{\rm B} T_{\rm C}$,\cite{gruner} which is 
19~meV from $T_{\rm C}$=63~K.   Hence, the observed onset 
of 10~meV is about twice smaller, and $\omega_{\rm CM}$ 
about three times larger, than the MF gap.

Note that the wide MF gap with the $\delta$ function-like 
peaks in DOS, predicted at $T$=0 as shown in Fig.~6(b), 
is never realized experimentally with real materials, even 
if $\sigma(\omega)$ is measured at much lower $T$ than 8~K.   
This is due to broadening casued by various extrinsic factors, 
such as crystal imperfection, present in real materials.    
Hence, the observed smaller transport gap than 
$\omega_{\rm CM}$ should be a result of a combination of 
such broadenings and the thermal fluctuation of the 
$f$ electrons discussed above.    
In any case, the model clearly shows that the evolution of 
the transport gap is not necessarily synchronous with that 
of the peak in $\sigma(\omega)$.    In particular, even when 
the transport gap is already suppressed, a peak in 
$\sigma(\omega)$ may still remain, if the peak in DOS still 
exists.   Such a situation may correspond to the observed 
$\sigma(\omega)$ at 14~GPa and 8~K: 
namely, although the Drude component is dominant and there 
is no clear energy gap, a Lorentz component is still 
present, as already discussed.   The presence of a peak in 
$\sigma(\omega)$ does not necessarily indicates the 
presence of a gap in the total DOS.

External pressure decreases the Pr-P distance, which should 
apparently increase $\Delta_0$, and should also increase $J$ 
since the $c$ 
electrons are from the P 3$p$ state.\cite{harima1,harima2}   
In addition, the pressure also decreases the Pr-Pr distance, 
which should increase the intersite coupling $K$.  Therefore, 
external pressure should affect all of $J$, $\Delta_0$ and 
$K$, so it is not obvious how the effective CF, 
$\Delta = \Delta_0 - J + \frac{3}{4}K$, changes with pressure.  
However, it has been suggested that the pressure should in 
fact increase $\Delta$,\cite{shiina2} based on comparison 
between the predicted phase diagram [Fig.~6(a)] and the 
$\rho(T)$ data under pressure.   This also applies well to 
$\sigma(\omega)$ data, since the $T$ and $P$ variations of 
$\sigma(\omega)$ show very good correspondence with those 
of $\rho(T)$, as already discussed.   
Note that the phase boundary near the vertical axis in 
Fig.~6(a) is predicted to be of first order.\cite{shiina2}   
This seems consistent with the $R(\omega)$ and $\sigma(\omega)$ 
data at 14~GPa since the changes (increases) in 
$R_{\rm d}(\omega)$ and $\sigma(\omega)$ from 30 to 8~K are 
much larger than the gradual changes from 60 to 30~K.  Based 
on these considerations, suggested lines corresponding to 
$P$=0, 10, and 14~GPa are schematically indicated in Fig.~6(a).   
From the theory, it is expected that the gap excitation 
peak should shift to lower energy with increasing $\Delta$, 
and hence with increasing $P$.\cite{shiina2}   
This is in contrast to 
the observed, only slight decrease of $\omega_{\rm CM}$ 
with $P$.    This will be discussed again later.

\subsection{Comparison with other CDW systems}  
Optical responses of more conventional CDW states driven 
by electron-lattice coupling have been studied 
extensively.\cite{gruner}    For example, $\sigma(\omega)$ 
of a series of compounds $R$Te$_3$ ($R$: rare earths) have 
been studied recently.\cite{RTe2,RTe3,RTe4}    A peak was 
commonly observed in their $\sigma(\omega)$ spectra, whose 
energy was identified as the CDW gap magnitude.   The peak 
shifted to lower energy, and hence the CDW gap decreased, 
with lattice contraction given by different 
$R$ elements\cite{RTe1,RTe2} and also by external 
pressure.\cite{RTe2,RTe3,RTe4} 
In contrast, as already mentioned, the gap excitation peak of 
PrRu$_4$P$_{12}$ decreases only slightly with $P$.   
The reason for this behavior is unclear at the present, and 
it may point to the unconventional nature of CDW state in 
this compound.  
The almost unshifted gap excitation peak and the progressive 
decrease of its spectral weight with $P$ may suggest a 
possibility that, while the gap magnitude remains unchanged, 
the fraction of FS that is gapped may decrease with $P$.   
Such a concept has been applied to analyze the optical 
data of the $R$Te$_3$ compounds.\cite{RTe1,RTe2,RTe3,RTe4}    
In the case of $R$Te$_3$ with layered crystal structure, 
the CDW had a strong 2D character, driven by a 2D nesting 
of FS.  As a result, the FS nesting was not complete and 
only a part of FS was gapped.      
In contrast, the nesting in PrRu$_4$P$_{12}$ is of 3D 
character\cite{harima1,harima2} as already mentioned, and 
the energy gap at $P$=0 is a full gap over the entire FS.   
Therefore, the above mentioned situation, namely the gap 
survives at some portion of the FS while it closes at 
other portion, seems unlikely for PrRu$_4$P$_{12}$.

Finally, it is also unlikely that the suppression of CDW state 
in PrRu$_4$P$_{12}$ is driven by a change of FS topology by 
pressure.     
In the pressure range of this work, PrRu$_4$P$_{12}$ does not 
undergo a structural phase transition and its crystal symmetry 
is unchanged with pressure.       
Although other parameters such as $\Delta_0$, $J$ and $K$ may 
change with pressure, the FS topology itself and the tendency 
for nesting should basically remain unchanged.\cite{harima3}  
This may be also different from the cases of 1D and 2D CDW 
compounds, where the low dimensionality may play an additional 
role for the external pressure to affect the FS topology.

\section{Conclusion}
The $\sigma(\omega)$ spectrum of PrRu$_4$P$_{12}$ has been 
measured under high pressure to 14~GPa and at low 
temperatures to 8~K, over a photon energy range of 12~meV-1.1~eV.   
The energy gap in $\sigma(\omega)$ associated with the 
insulating, unconventional CDW state at ambient pressure 
is progressively filled in with increasing pressure.    
At 14~GPa and 8~K, $\sigma(\omega)$ exhibits a strong Drude 
component due to free carriers, and the energy gap is almost 
absent.    Our finding confirms the appearance of metallic 
state that had been suggested by $\rho(T)$ data under pressure.    
Effects of pressure on the $c$-$f$ hybridization and the 
unconventional CDW state has been discussed based on our 
data as well as microscopic theories for PrRu$_4$P$_{12}$.

\begin{acknowledgements}
The experiments at SPring-8 have been made under the 
approval by JASRI (2009A0089 through 2011B0089).  
Financial support from MEXT (``Heavy Electron'' 
21102512-A01) is acknowledged.  
U2A beamline is supported by NSF (EAR 10-43050, COMPRES), 
and DOE/NNSA (DE-FC03-03N00144,CDAC). 
NSLS is supported by the DOE/BES (DE-AC02-98CH10886).
H. O. would like to thank R. Shiina, H. Harima, and 
Y. Aoki for useful discussions.  
\end{acknowledgements}



\begin{thebibliography}{99}
\bibitem{footnote3}  
The deviation between the $R_{\rm d}(\omega)$ actually measured 
in DAC and that calculated from the $R_0(\omega)$ was within 
20~\% in the spectral range above 50~meV.  
Below 50~meV, in contrast, the former decreased with decreasing 
photon energy due to diffraction, to about 60~\% magnitude of 
the latter at 20~meV.   
Nevertheless, the relative $T$ variation actually measured 
in DAC agreed reasonably well with the calculated ones from 
the $T$ variations of $R_0(\omega)$, as previously 
demonstrated.\cite{CeRu4Sb12}
\bibitem{footnote4}
The experimentally observed singlet-triplet splitting is 
$\sim$ 8~meV above $T_{\rm C}$, and 3~meV and 10~meV for 
the triplet and singlet ground states at 5~K, 
respectively.\cite{iwasa1,iwasa2,ogita}
\end{thebibliography}
\end{document}